\documentclass[prb,preprint,showpacs]{revtex4-1}

\usepackage{epstopdf}
\usepackage{graphicx}  % needed for figures
\usepackage{dcolumn}   % needed for some tables
\usepackage{bm}        % for math
\usepackage{amssymb}
\usepackage[english]{babel}
\usepackage{lineno}
\begin{document}

%\preprint{APS/123-QED}

\title{High Pressure Structural Investigation on Lead-Free Piezoelectric $0.5Ba(Ti_{0.8}Zr_{0.2})O_3 - 0.5(Ba_{0.7}Ca_{0.3})TiO_3$}

\author{Anshuman Mondal}
\affiliation{Department of Physical Sciences, Indian Institute of Science Education and Research Kolkata, Mohanpur Campus, Mohanpur 741246, Nadia, West Bengal, India.}

\author{Rajeev Ranjan}
\affiliation{Department of Materials Engineering, Indian Institute of Science, Bengaluru, Karnataka 560012, India.}

\author{Kumar Brajesh}
\affiliation{Department of Material science and Engineering, IIT Kanpur,  560012, India.}

\author{Pinku Saha}
\affiliation{Department of Physical Sciences, Indian Institute of Science Education and Research Kolkata, Mohanpur Campus, Mohanpur 741246, Nadia, West Bengal, India.}

\author{Bishnupada Ghosh}
\affiliation{Department of Physical Sciences, Indian Institute of Science Education and Research Kolkata, Mohanpur Campus, Mohanpur 741246, Nadia, West Bengal, India.}

\author{Mrinmay Sahu}
\affiliation{Department of Physical Sciences, Indian Institute of Science Education and Research Kolkata, Mohanpur Campus, Mohanpur 741246, Nadia, West Bengal, India.}

\author{Goutam Dev Mukherjee}
\email [Corresponding author:]{goutamdev@iiserkol.ac.in}
\affiliation{Department of Physical Sciences, Indian Institute of Science Education and Research Kolkata, Mohanpur Campus, Mohanpur 741246, Nadia, West Bengal, India.}

\begin{abstract}

The solid solution $0.5Ba(Ti_{0.8}Zr_{0.2})O_3 - 0.5(Ba_{0.7}Ca_{0.3})TiO_3$ (BCZT) has become a promising member of the lead-free piezoelectric materials because of its exceptionally high piezoelectric properties. In this study, we focus on studying pressure-dependent Raman spectroscopy, powder x-ray diffraction and dielectric constant measurements on BCZT. The data show several structural transitions are present, where the system from ambient mixed phase (tetragonal, {\it {P4mm}}+ orthorhombic {\it {Amm2}}) transforms into single phase ({\it {P4mm}}) at 0.26 GPa, then converts into cubic phase ({\it {Pm3m}}) at 4.7 GPa followed by another possible structural re-ordering around 10 GPa. Although there have been a lot of unanimity with the ambient crystallographic state of BCZT, our analysis justifies the presence of an intermediate orthorhombic phase in the Morphological Phase Boundary (MPB) of BCZT phase diagram. The transformation tetragonal to cubic is indicated by the Raman mode softening, unit cell volume change and the $(Ti/Zr)O_6$ octahedra distortion, which coincides with the well-known ferroelectric-paraelectric transition of the system. The sudden drop in the dielectric constant value at 4.7 GPa also confirms the loss of ferroelectric nature of the BCZT ceramic.

\end{abstract}

\date{\today}
\maketitle

%\tableofcontents

\section{Introduction}

During last few decades, to avoid the hazardous effects of lead (Pb), there have been ongoing efforts to replace the commercially used lead-zirconate-titanate (PZT) based piezoelectric materials with the lead-free materials having high piezo-response \cite{cross2004materials, rodel2009perspective, coondoo2013synthesis, porta2011effects, bai2013optimized}. Among them,  $0.5Ba(Ti_{0.8}Zr_{0.2})O_3 - 0.5(Ba_{0.7}Ca_{0.3})TiO_3$ (BCZT) has drawn significant attention because of its very high piezoelectric properties ($\sim$ 600 pC/N) \cite{liu2009large, damjanovic2012elastic} in comparison with that of PZT (500-600 pC/N). The main reason lies within the phase diagram of solid solution BCZT having a Morphological Phase Boundary (MPB) that includes a triple point at the intersection of paraelectric cubic, ferroelectric rhombohedral and tetragonal phases \cite{liu2009large}. Also it is reported that the piezoelectric properties are maximized at the tetragonal-rhombohedral phase boundary \cite{liu2009large}. But there has been an ongoing debate on the crystallographic state of the BCZT composition at ambient conditions. Studies by Keeble et al.\cite{keeble2013revised} \& Tian et al.\cite{tian2014polymorphic} have proposed that there exists an intermediate orthorhombic (O) phase in between rhombohedral (R) and tetragonal (T) phases making the "polarization rotation" possible between the two crystallographic groups. Furthermore, experimental studies by Zhang et al.\cite{zhang2014phase} and Acosta et al.\cite{acosta2014relationship} confirm that instead of the MPB (T-R phase boundary), the T-O phase boundary is responsible for the large piezo-response of the BCZT ceramics. It is also shown that a mere 2\% of Zr, Sn or Hf substitution in parent $BaTiO_3$ can stabilize a co-existence of orthorhombic and tetragonal phases at room temperature \cite{kalyani2014orthorhombic}. Although presence of this orthorhombic phase is countered and also argued as an anomaly that arises due to the adaptive diffraction of tetragonal/rhombohedral nanodomains\cite{gao2011microstructure,gao2014symmetry,bjornetun2013structure,ehmke2014resolving}. Brajesh et al.\cite{brajesh2015relaxor,brajesh2016structural} has performed careful structural and dielectric studies on BCZT and concluded in favour of three-phase co-existing model (P4mm+Amm2+R3m) to explain its giant piezo-response and relaxor ferroelectric nature. But it may give rise to a scenario of violating Gibb's phase rule if the orthorhombic phase is present all the way up to the critical point of the phase diagram. The study by Yang et al.\cite{yang2016mechanisms} explains the largest piezoelectricity as well as the largest elastic softening of BCZT at the T-O phase boundary by the systematic energy barrier (EB) calculations that measures the degree of polarization anisotropy. So, one needs careful investigation on the ambient crystallographic state of the BCZT ceramic.

BCZT is one of the most well studied systems. Microstructure basis for strong piezoelectricity in BCZT via transmission electron microscopy (TEM) is shown by Gao et al. \cite{gao2011microstructure}. Fan et al. have carried out an {\it{in situ}} TEM study using unipolar electric field and shown that ferroelectric domains are disrupted by defect states \cite{fan2017domain}. The effect of grain size on the electrical properties and temperature dependance of variation in dielectric constant are shown by Hao et al. \cite{hao2012correlation}. Temperature induced anomalies in the dielectric, piezoelectric and elastic coefficients and Raman spectroscopy of ceramic BCZT are reported by Damjanovic et al. \cite{damjanovic2012elastic}. Piezoelectric property measurements by Piezoresponse Force Microscopy (PFM) mode of Scanning probe microscopy (SPM) technique is reported by Coondoo et al. \cite{coondoo2013synthesis}, whereas piezo-properties on pulsed laser-deposited thin films are reported by Piorra et al. \cite{piorra2011piezoelectric}. Based on all these available literature, it can be concluded that the piezoelectric properties of lead-free BCZT system is comparable to the PZT. But after the Curie temperature ($T_c$) at $97^o$ C, all studies have shown a drop in the temperature dependent dielectric constant values and have concluded this to loss of ferroelectric nature of the sample following a phase transition into cubic phase. The curie temperature ($T_c$) has also been modified by grain size changing or even by doping.

Although there are several temperature dependent studies, to the best of our knowledge, BCZT certainly lacks the pressure dependent measurements. BCZT, which is a solid solution of BZT-BCT, hold all the properties of its parent material $BaTiO_3$ along with the effect of doping on it. Therefore, the present manuscript focuses on high-pressure Raman spectroscopy, x-ray diffraction and dielectric measurements of BCZT to study phase transitions, which modify its dielectric properties and piezoelectric nature.

\section{Experiment}
Polycrystalline powder sample $0.5Ba(Ti_{0.8}Zr_{0.2})O_3 - 0.5(Ba_{0.7}Ca_{0.3})TiO_3$ (or BCZT) is prepared via conventional solid-state reaction method \cite{brajesh2016structural}. High-pressure x-ray diffraction (XRD) and Raman spectroscopic measurements are performed using diamond-anvil cell (DAC) from EasyLab Co. (UK) with a culet size of 300 $\mu$m. To hold the sample in between two opposing diamonds, a pre-indented metal gasket (indented thickness $\sim$50 $\mu$m) is used with a 100 $\mu$m see-through hole at the centre of indentation. Methanol-ethanol mixture (4:1) is used as the pressure-transmitting medium (PTM) to maintain the hydrostatic condition. For the Raman-spectroscopic measurements, Monovista CRS+ spectrometer is used along with an infinitely corrected long-working distance 20X objective. Pressure is determined by observing the shift of ruby fluorescence peak with ruby grains loaded alongside the sample inside DAC \cite{mao1986calibration}. Sample is excited using a 532 nm green laser and spectra are collected using a grating of 1500 grooves/mm in the back-scattering geometry with resolution of 1.2 $cm^{-1}$.

High-Pressure XRD measurements are performed at the XPRESS beamline in the Elettra Synchrotron Source, Italy using a monochromatic wavelength of 0.4957 $\AA$. For XRD, a small amount of Ag powder is mixed along with the sample and pressure is calculated using the well-known Birch-Murnaghan equation of state of Ag \cite{dewaele2008compression}. The incident x-ray is collimated to 30 $\mu$m and a MAR-3450 image plate detector is used to detect the diffracted x-rays, aligned normal to the beam. The sample to detector distance is calibrated using $LaB_6$ and all the diffracted patterns are integrated to intensity vs 2$\theta$ profiles using FIT2D software \cite{hammersley1996two} and then further analyzed using CRYSFIRE \cite{Crysfire} and GSAS \cite{toby2001expgui}.

High-pressure dielectric constant measurements are carried out up to 5.5 GPa using the large-volume Toroid-anvil (TA) cell and a 300-ton hydraulic press. The TA apparatus is calibrated using Bi I-II and Yb hcp-bcc transitions at 2.65 GPa and 4 GPa respectively \cite{achary2002new}. The sample assembly is prepared and the frequency dependent capacitance is measured as described by Jana et al. \cite{jana2018pressure}. The dielectric constant (K) is calculated using the expression, $C = K\epsilon_0(A/d)$, where $\epsilon_0$ is the permittivity of the free space, C is the capacitance, A is the area of the copper plate and d is the thickness of the sample pellet.

\section{Results and discussions}
%%\subsection{X-ray diffraction studies}

BCZT can be thought as Ca and Zr doping of the parent sample $BaTiO_3$, making a solid solution out of it. The sample is characterized by XRD measurements. Fig. 1(a) shows the ambient XRD pattern. Indexing of the sample at ambient condition leads to a mixed phase of tetragonal and orthorhombic crystal structures supporting earlier reports\cite{yang2016mechanisms,kalyani2014orthorhombic}. Using the initial atom positions given by Brajesh et al.\cite{brajesh2016structural} for the two phases, full structural Rietveld refinement has been carried out using the GSAS \cite{toby2001expgui} software. In this refinement process, only the atom positions corresponding to Ti/Zr atom are refined and oxygen being a lighter atom is exempted from this refinement process\cite{brajesh2015relaxor}. The lattice parameters obtained from the best-fit are - (i) for tetragonal phase: a = 3.99906(9) \AA, c = 4.0208(1) \AA, vol = 64.303(3) $\AA^3$ and (ii) for orthorhombic phase: a = 3.9980(1) \AA, b = 5.6775(2) \AA, c = 5.6670(1) \AA, vol = 128.635(7) $\AA^3$. In Fig. 2, the singlet nature of the $2\theta = 12.31^o$ peak (111) and the doublet nature of the peak around $2\theta = 14.2^o$ [(002)/(200)] in the ambient XRD pattern are implying a tetragonal (P4mm) global structure \cite{brajesh2015relaxor}. The asymmetric doublet nature of the peak around $2\theta = 20.14^o$ [(040)/(222)] in Fig. 2(c) and the peak at $2\theta = 22.54^o$ [(133)/(311)] (Fig. 2(d)), which can only be fitted with the orthorhombic phase, indicate the perovskite structure to be orthorhombic in nature too. With these two phases, we have an excellent fit that certainly sets aside the question of any anomaly that may have come due to the adaptive diffraction by the P4mm-R3m nanodomains as reported by Gao et al.\cite{gao2014symmetry} earlier. At ambient conditions, the presence of both the tetragonal and orthorhombic phases are justified.

Ambient and high pressure Raman spectroscopy measurements are performed on BCZT to explore the microscopic changes that occur due to the volume compression. The ambient Raman spectrum is shown in Fig 3(a). The pattern shows broad Raman modes and looks similar to that of parent $BaTiO_3$ \cite{venkateswaran1998high}. To determine the Raman mode frequency positions, we have performed the Lorentz fitting of each peak. Most of the modes are indexed following $BaTiO_3$ Raman mode assignments carried out by Venkateswaran et al.\cite{venkateswaran1998high}. The most important feature of this kind of systems is the presence of several anharmonic modes in the low frequency range. The dip at 130 $cm^{-1}$ is referred as anharmonic coupling mode that comes due to the interference between the three $A_1(TO)$ modes. All the modes with their corresponding frequency positions are listed in Table 1. We also have observed two new peaks $M_1$ (88 $cm^{-1}$) and $M_2$ (97 $cm^{-1}$), which could not be identified with $BaTiO_3$ modes. The low frequency modes are due to vibration of heavy atoms. Therefore, these extra peaks can be related to the substitution of Ca and/or Zr ions and the defects that might have occurred due to these substitutions.
\begin{table}[h]
	\centering
	\caption{\textbf{Ambient Raman mode frequency of BCZT}}
	\begin{tabular}{|c| c|}
		\hline
		&\\
		\textbf{Mode Symmetry} &  \textbf{Mode Frequency($cm^{-1}$})\\
		&\\
		\hline
		$M_1$  & 88  \\
		$M_2$ & 97 \\
		$A_1(TO_1)$ & 149\\
		$E(TO_1),E(LO)$ & 176\\
		$A_1(LO_1)$ & 205\\
		$A_1(TO_2)$ & 245\\
		$B_1 + E(TO_2+LO)$ & 293\\
		$E(TO_3)$ & 525\\
		$A_1(TO_3)$ &  543\\
		$E(LO)$ & 737\\
		$A_1(LO_2)$ &  809\\
		\hline
	\end{tabular}
\end{table}
\noindent Raman spectra evolution with pressure is shown in Fig. 3(b). With pressure, the dip around 130 $cm^{-1}$ reduces, and similar observation has also been found by Basu et al. \cite{basu2015pressure}. The reason behind this can be the increase of anharmonic contributions of all the low frequency modes, that is evident from Fig. 3(b). Whereas, all the modes broaden and their frequencies increase with pressure, anomalous behaviour is observed for the $A_1(TO_2)$ and $E(TO_3)$ modes.  The $A_1(TO_2)$ mode softens with increase of pressure up to 5.3 GPa, then starts increasing (see Fig 4). $E(TO_3)$ mode also softens with a slope change above 5 GPa. The high frequency modes can be related to light atom vibrations such as, Ti-O/Zr-O bond vibrations. Therefore, this anomalous behaviour of $E(TO_3)$ mode is certainly due to the $(Ti/Zr)O_6$ octahedral distortion in the system that tends to gain more symmetry with applied pressure. The sharp kink at 293 $cm^{-1}$ looses its sharp nature  beyond 4.4 GPa. The asymmetric peak around 525 $cm^{-1}$ loses its intensity with pressure and shows a second peak on its asymmetric tail (Fig 3(b)) for pressures above 3.5 GPa. Above 4.4 GPa, the new peak tends to increase, exceeding the old one in intensity. The pressure induced modification leads to the loss of the $A_1(LO_1)$ mode beyond 4.4 GPa. Two new modes are also observed from 4.4 GPa onwards at 124 $cm^{-1}$ and 632 $cm^{-1}$ referred as $\omega_1$ and $\omega_2$ respectively (see Fig 3(b)). In the low frequency range, appearance of new modes can be related to heavy atom (Ba/Ca) vibrations\cite{jana2018pressure}.
Softening of low frequency Raman modes are associated with decrease in polarization in the sample leading to change in ferroelectric nature \cite{venkateswaran1998high}. Softening of modes, appearance of two new modes ($\omega_1$ and $\omega_2$), disappearance of $A_1(LO_1)$ mode, loss of sharp nature of the peak at 293 $cm^{-1}$ and decrease in intensity of the 525 $cm^{-1}$ mode are the characteristics of change in crystal structure.   
The weak peak near 886 $cm^{-1}$ can be seen throughout all pressure plots except the ambient spectra. The position of this mode changes at very high rate (3.26 $cm^{-1}$/GPa) and can be related to the C-C stretching of ethanol-methanol pressure transmitting medium. Beyond 5 GPa, with increase in pressure we have observed several other additional interesting features. It includes the loss of $A_1(TO_3)$ mode at 10.3 GPa and several new peaks appear at 333 $cm^{-1}$ ($\omega_3$), 809 $cm^{-1}$ ($\omega_4$) and 418 $cm^{-1}$ ($\omega_5$) from 5.9, 9.6 and 12.5 GPa pressure points respectively (see Fig 3(b)). With pressure, the frequency of all these new modes increase up to the highest pressure point of our study. The $\omega_3$ mode initially increases up to about 11.7 GPa followed by a sharp drop (Fig 4). Softening of certain Raman modes and appearance of several new modes indicate to certain structural re-ordering. Softening of $A_1(TO_2)$ and $E(TO_3)$ modes indicate to loss of ferroelectricity. Therefore application of pressure pushes the system into a more symmetric paraelectric phase. The presence of Raman spectra beyond this transition in the paraelectric phase can be explained by order-disorder model in which the Ti/Zr atoms are not sitting in the exact body-centered position of the octahedra. It is slightly distorted along the body diagonal which breaks the inversion symmetry allowing the Raman activity \cite{venkateswaran1998high, sood1995phonon}. However, direct structural determination is not possible using Raman spectroscopy; but it is more likely to present the instantaneous changes in atomic positions. X-ray diffraction studies are required for further understanding.

To observe a direct change in the ferroelectric properties of BCZT, we have performed high-pressure dielectric constant measurements. Fig. 5(a) shows the variation in dielectric constant(K) and dielectric loss($\delta$) values from 0.35 GPa to up to 5.41 GPa for some selected frequency values (50 Hz, 100 Hz, 500 Hz, 1 KHz, 5 KHz, 10 KHz, 50 KHz, 100KHz, 500 KHz, 1 MHz). At each frequency, 10 consecutive measurements of capacitance and loss are taken and the mean value is determined. The value of the dielectric constant at 0.35 GPa is matching with the reported result \cite{scarisoreanu2015high}. At all pressure points, K shows an exponential-like decrease with frequency. In addition at a constant frequency (mainly at low frequency) K  value decreases with increase in pressure and above 4.69 GPa it seems to saturate. Now, if we observe the dielectric constant value at 0.8 GPa at 50 Hz frequency
(shown in Fig. 5(a)), it is around 1875. Such a large value of dielectric constant can come from both the contribution of intrinsic and extrinsic factors. These extrinsic factors include contributions due to the Maxwell-Wagner effect and also the effect of grain boundaries in such a polycrystalline system. To properly understand this, we have plotted the variation of dielectric constant and dielectric loss value with frequency at 6.82 GPa pressure (see Fig. 5(b)). Here the dielectric constant shows a drop and this drop matches with the peak of dielectric loss in between 500-1000 Hz indicating a Debye-like relaxor behavior arising due to the Maxwell-Wagner effect \cite{jana2016high}. Strong frequency dependence in dielectric constant can be modeled in terms of an equivalent circuit consisting of two parallel R-C circuits connected in series, where one of the capacitance comes from the extrinsic effects and another corresponds to bulk intrinsic effects\cite{jana2016high}. At high frequencies, above the peak frequency of the dielectric loss, the capacitance arising due to extrinsic effects get shorted, and intrinsic bulk behavior is observed. Therefore to study the pressure evolution of dielectric constant (see in Fig. 5(c)), we have considered the 100 KHz frequency i.e. a high-frequency range data. In Fig. 5(c), we have shown the pressure dependence of dielectric constant K and dielectric loss, at frequency 100 KHz. In this frequency, at 0.35 GPa, the K value is calculated to be 61. As pressure increases, K almost remain constant up to 0.8 GPa, then rapidly rises to 96 at 4.37 GPa pressure, followed by a sharp drop at about 4.7 GPa signifies the loss of the ferroelectric nature of BCZT. Then afterward, it shows a flat nature with almost no change in value.

To understand the direct structural changes, ambient and high pressure x-ray diffraction studies are performed. While indexing the XRD patterns at 0.26 GPa, we get a better match with only tetragonal phase. The hkl-values have been assigned following Coondoo et. al \cite{coondoo2013synthesis} and the Rietveld refinement provides us a best-fit that is shown in Fig. 1(b). If we examine the XRD patterns from 4.8 GPa onwards, then we can see that the doublet peaks (shown in Fig.6) in the tetragonal phase (Space Group: P4mm) merge to form a single Bragg reflection, which implies a phase transition happening in between 4.4 GPa and 4.8 GPa. Fig.6 illustrates this in detail. The indexing of the XRD data at 4.8 GPa gives a cubic structure with Pm3m space group and after Rietveld refinement (shown in Fig. 1(c)), the best-fit gives us the value of lattice parameter, a = 3.9640(1) $\AA$ with cell volume = 62.288(3) $\AA^3$. We have plotted $c/a$ ratio with pressure for tetragonal phase and shown in Fig 7. The value of $c/a$ ratio changes from 1.005 to 1 at 4.8 GPa, indicating a transformation to cubic structure. The system remains in this configuration up to the highest pressure value of our study, 20.5 GPa. After the Rietveld refinement, the unit cell volume obtained from each XRD pattern of different pressure points have been used in the EoSFit program for the BM-EOS fitting \cite{angel2014eosfit7c}. The volume data as a function of pressure both in the tetragonal and cubic phases are fitted to the 3rd order Birch-Murnaghan equation of state (BM-EOS) represented by
\begin{equation}
	P(V)=\frac{3 B_{0}}{2}\left[\left(\frac{V_{0}}{V}\right)^{7 / 3}-\left(\frac{V_{0}}{V}\right)^{5 / 3}\right] \times\left\{1+\frac{3}{4}\left(B^{\prime}-4\right)\left[\left(\frac{V_{0}}{V}\right)^{2 / 3}-1\right]\right\}
\end{equation}
\noindent
where $V_0$ is the initial volume (in ambient), $B_0$ is the bulk modulus, and $B'$ is the first pressure derivative of bulk modulus. From the fitting (see Fig. 8(a) and 8(b)), we get the best-fit values along with $\chi^2$-value of fit. The values obtained from the best-fit lines are : (a) for tetragonal phase: $V_0$ = 64.3(1) $\AA^3$, $B_0$ = 149(3) GPa; (b) for cubic phase: $V_0$ = 64.2(1) $\AA^3$, $B_0$ = 145(1) GPa. The inverse of the bulk modulus gives us the compressibility. From Fig. 8(a) and 8(b), we can say that although there is a very little difference in the Bulk modulus values in the two phases, still the compressibility in the tetragonal phase is less compared to the Cubic phase. The slight difference in the bulk moduli between the above two phases signify that other than the effect of pressure on the lattice parameters of the system, some other factor is more responsible for this transition. From Raman Spectroscopy measurements this part can be identified to be the octahedral distortion. To understand this, using the final refinement results, a 3-D polyhedron model of BCZT has been generated for each pressure values using the VESTA software \cite{momma2011vesta}. Now, we wish to understand how the $(Ti/Zr)O_6$ octahedra is changing with pressure. The variation of polyhedral volume and the distortion index is shown in Fig. 8(c) The huge change in the polyhedral volume is clearly implying the tetragonal to cubic phase transition. Whereas in Fig. 8(d), we observe that the distortion index is increasing very rapidly from ambient to 5.5 GPa, after that it almost reaches to a saturation, then again after 10 GPa, the distortion index starts increasing, but the rate is very low. It can be summed up that with the application of pressure, the $(Ti/Zr)O_6$ octahedra is getting distorted very rapidly and trying to get into a more symmetric cubic structure, then after reaching there it stabilizes, for that we see the saturation. However further increase in pressure makes the system a bit unstable,  that is why a slight increasing nature of the distortion index after 10 GPa is observed. This slight change in the distortion index beyond 10 GPa might indicate a possible isostructural transition as proposed by Moriwake et al.\cite{moriwake2008isostructural} for the parent $BaTiO_3$. Appearance of new high frequency Raman modes ($\omega_4$ and $\omega_5$) also corroborate the above fact. Our high pressure work on this important piezoelectric material BCZT shows that proper characterization under strained condition is important to predict and understand its electrical properties, which may be useful in future applications.

\section{Conclusion}
In conclusion, high-pressure Raman spectroscopy, x-ray diffraction and AC dielectric measurements have been carried out for $BaTiO_3$ based lead-free piezoelectric sample $0.5Ba(Ti_{0.8}Zr_{0.2})O_3 - 0.5(Ba_{0.7}Ca_{0.3})TiO_3$. High-pressure XRD studies have indicated two structural phase transitions. One is at 0.26 GPa, where the system changes from mixed phase to only tetragonal phase. The initial flat nature of the graph of pressure dependent dielectric constant measurement was also indicative to this transition. Another one is at about 4.8 GPa, which corresponds to the tetragonal-cubic phase transition. The sudden change in the dielectric constant value at 4.7 GPa, corroborates the aforesaid result. High-pressure Raman spectroscopy results also compliment this finding. Softening of certain Raman modes and line shape changes in the Raman spectra confirms the ferroelectric to paraelectric transition. The presence of Raman modes even in the cubic phase is due to the distortion in the $(Ti/Zr)O_6$ octahedra. 

\section{Acknowledgments}
The authors gratefully acknowledge the Ministry of Earth Sciences, Government of India, for the financial support under the grant No. MoES/16/25/10-RDEAS to carry out this high pressure research work. A.M. also gratefully acknowledges the fellowship grant from the CSIR program by the Human Resource Development Group, Government of India.  The authors also gratefully acknowledge the financial support from the Department of Science and Technology, Government of India to visit XPRESS beamline in the ELETTRA Synchrotron light source under the Indo-Italian Executive Program of Scientific and Technological Cooperation.

%\section{References}

\newpage

\begin{figure}
	%\begin{center}
		\includegraphics{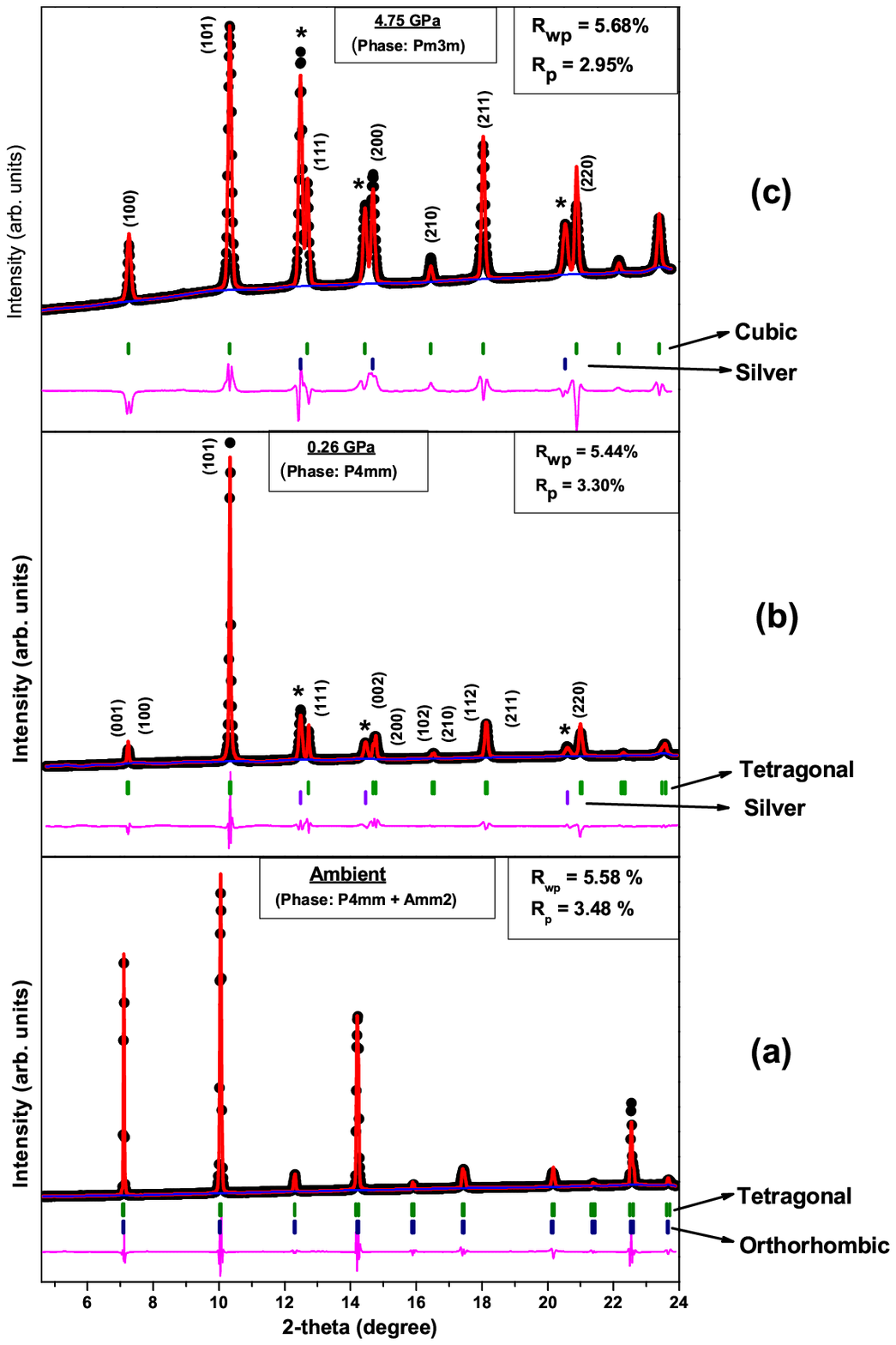}
		\caption{Rietveld refinement of BCZT XRD patterns at (a) ambient pressure (with mixed phase tetragonal and orthorhombic), (b) 0.26 GPa (with only tetragonal phase) and (c) 4.75 GPa (with cubic phase). The black dots represent the data points, the red line is the fitted line, and the pink line below represents the difference between the observed and the fitted intensities. '*' indicates the pressure-marker Silver(Ag). The goodness of fit parameters ($R_{wp}$ and $R_p$) is given in the boxes.}
	%\end{center}
\end{figure}

\begin{figure}
	%\begin{center}
		\includegraphics{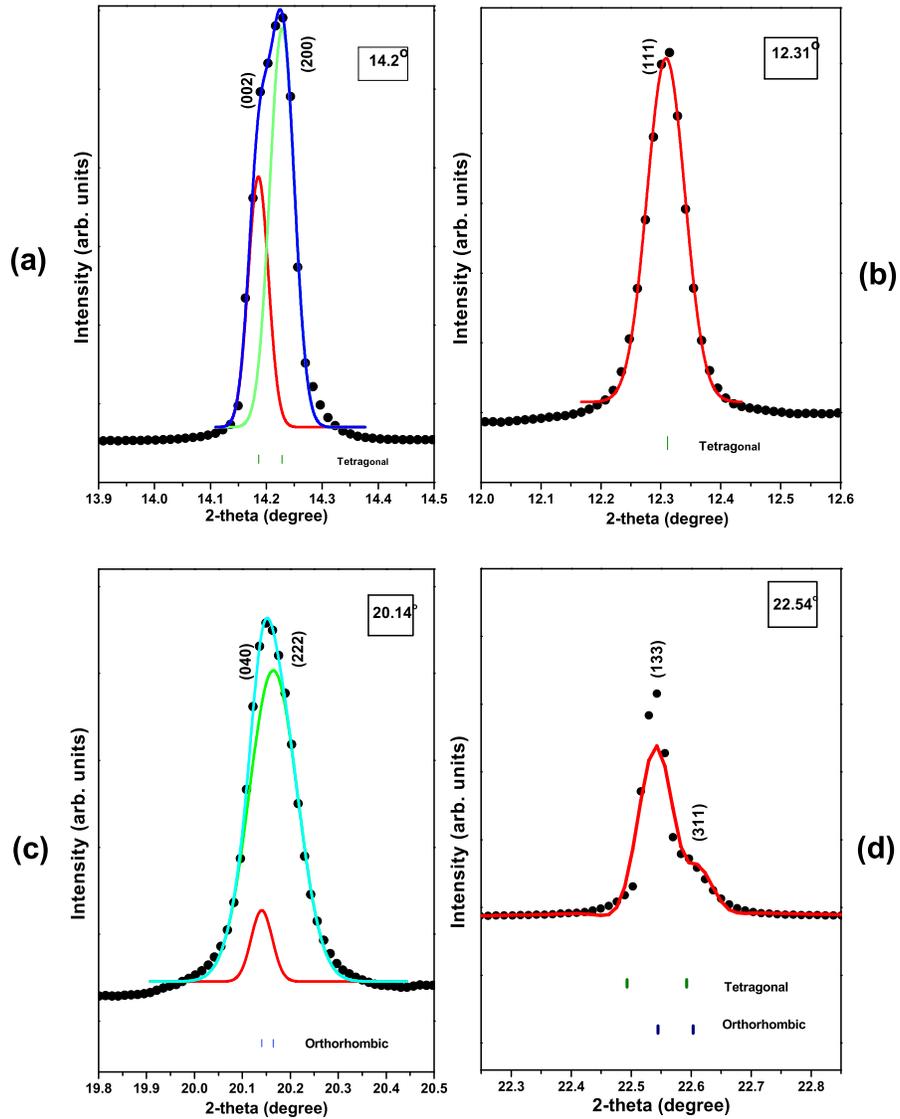}
		\caption{Fitting of Bragg peaks implying tetragonal and orthorhombic nature of ambient BCZT. (a) Doublet nature of $14.2^o$ peak and (b) singlet nature of $12.31^o$ peak imply the structure to be tetragonal. (c) The doublet nature of $20.14^o$ peak and (d) the $22.54^o$ peak is fitted best with the orthorhombic phase imply the presence of orthorhombic phase too.}
	%\end{center}
\end{figure}

 \begin{figure}
 	%\centering
 	%\includegraphics[width=0.8\textwidth]{fig3.eps}
	\includegraphics{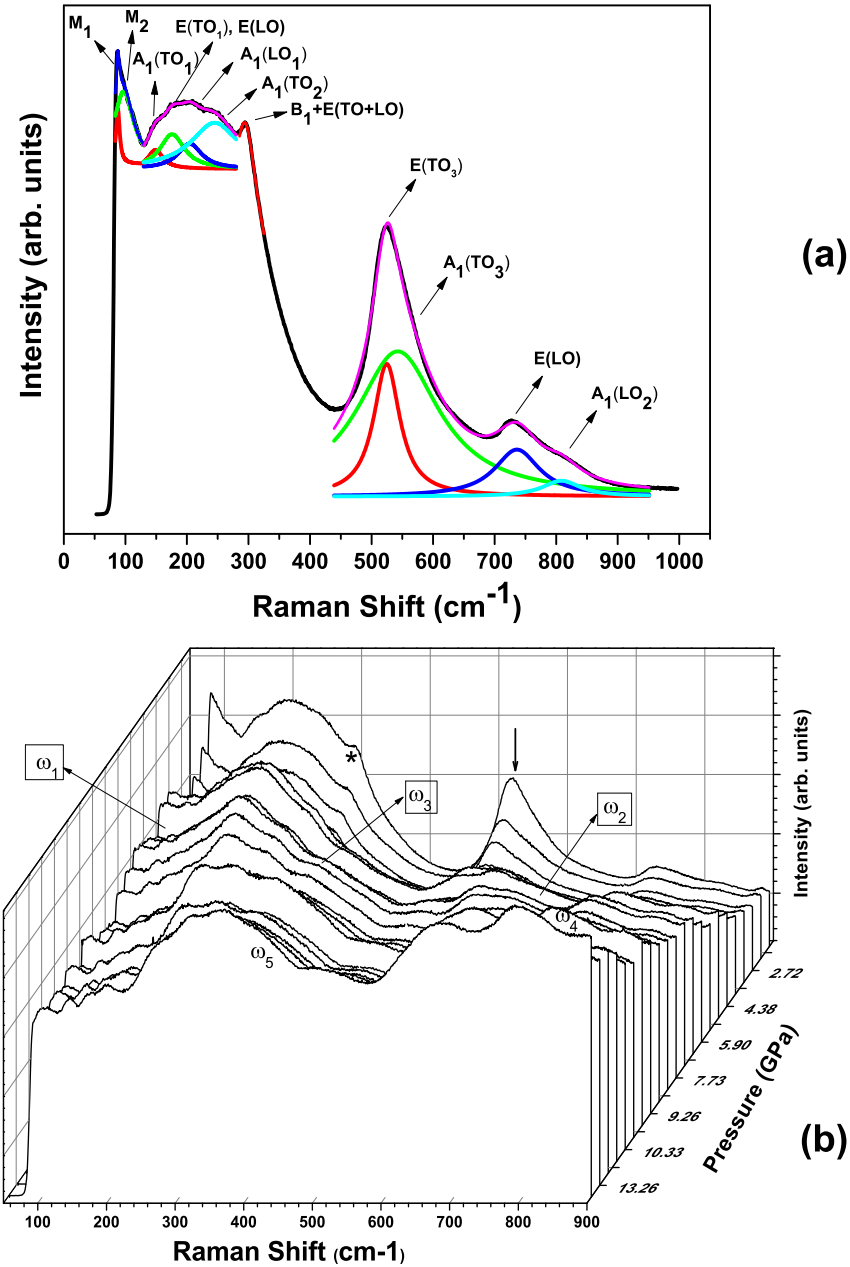}
 		\caption{(a)De-convolution of BCZT ambient Raman spectra and Raman mode assignment. (b) Evolution of Raman spectra starting from 1.36 GPa(top)  upto 13.88 GPa(bottom). The arrow indicate the $E(TO_3)$ mode which softens and reduces its intensity rapidly with pressure and finally shows a second peak from its asymmetric tail above 3.47 GPa. The sharp nature nature of the "*" peak also reduces. With pressure softening of modes, line-shape change in low frequency range and appearance of several new peaks ($\omega_1$, $\omega_2$, $\omega_3$, $\omega_4$ and $\omega_5$ respectively) indicate structural rearrangement.}
 \end{figure}

\begin{figure}
	%\begin{center}
		\includegraphics{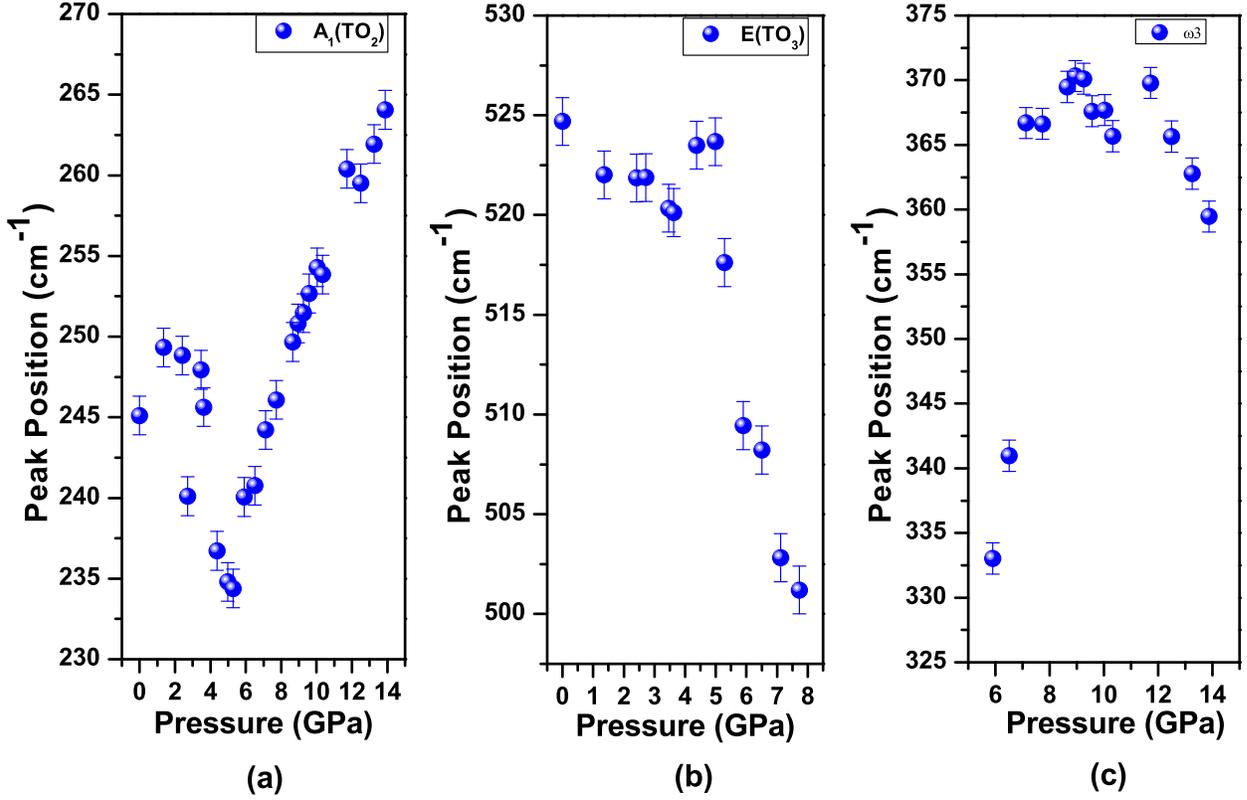}
		\caption{Evolution of $A_1(TO_2)$, $E(TO_3)$ and $\omega_3$ modes with pressure. (a) $A_1(TO_2)$ mode initially softens up to 5.29 GPa, then increases with further increase in pressure. (b) $E(TO_3)$ mode shows an overall softening kind of nature with pressure up to 8 GPa with a slope change at about 5 GPa. (c) The frequency position of the new peak $\omega_3$ initially increases, followed by a large drop above about 11 GPa implying some sort of structural re-ordering. }
	%\end{center}
\end{figure}

 \begin{figure}
% 	\centering
 	\includegraphics{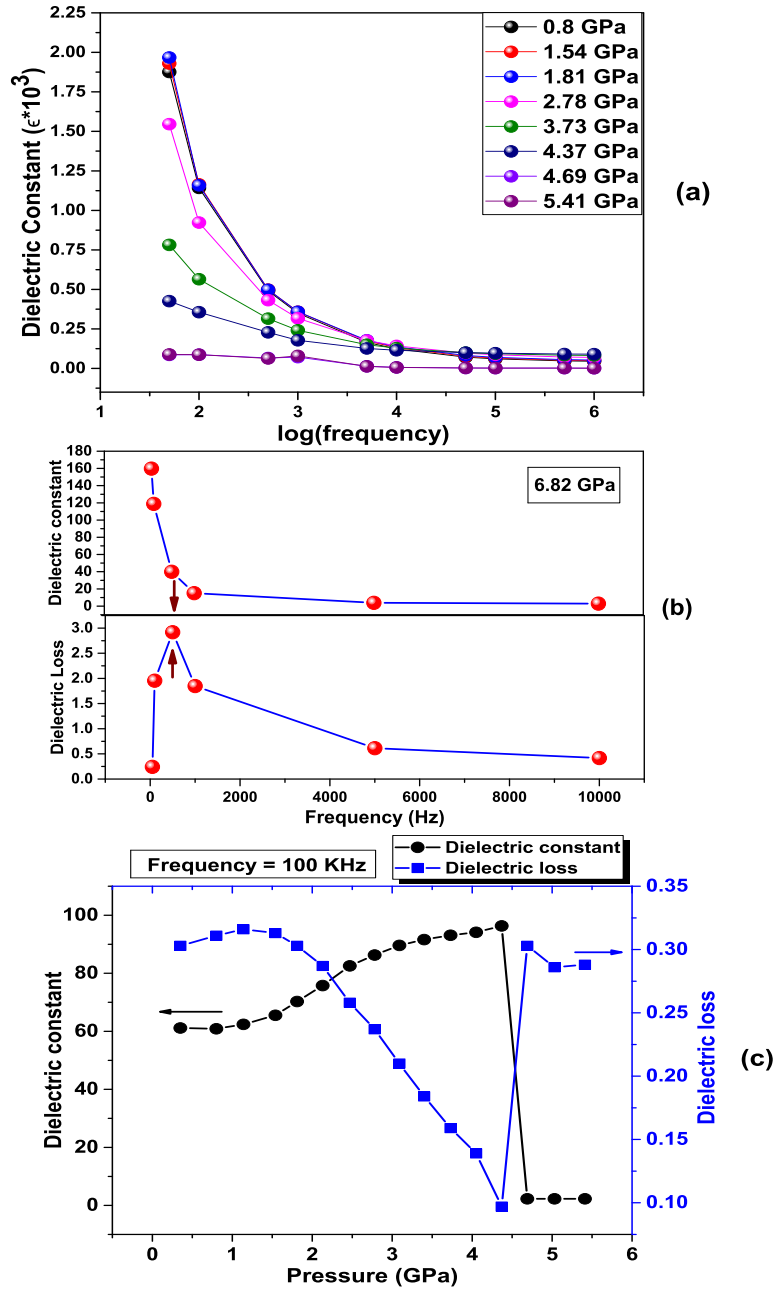} 	    
	\caption{(a)Frequency dependence of the dielectric constant value
			of BCZT at selected pressures showing exponential like decreasing nature. (b) Frequency dependence of dielectric constant and loss at 6.82 GPa pressure showing Debye-like relaxation behavior at high frequencies. (a) Pressure dependence of dielectric constant K (shown in dots) and dielectric loss (shown in squares) of BCZT at 100KHz. The sharp drop of dielectric constant value at 4.7 GPa indicating loss of ferroelectric nature.}
\end{figure}

\begin{figure}
	%\begin{center}
		\includegraphics{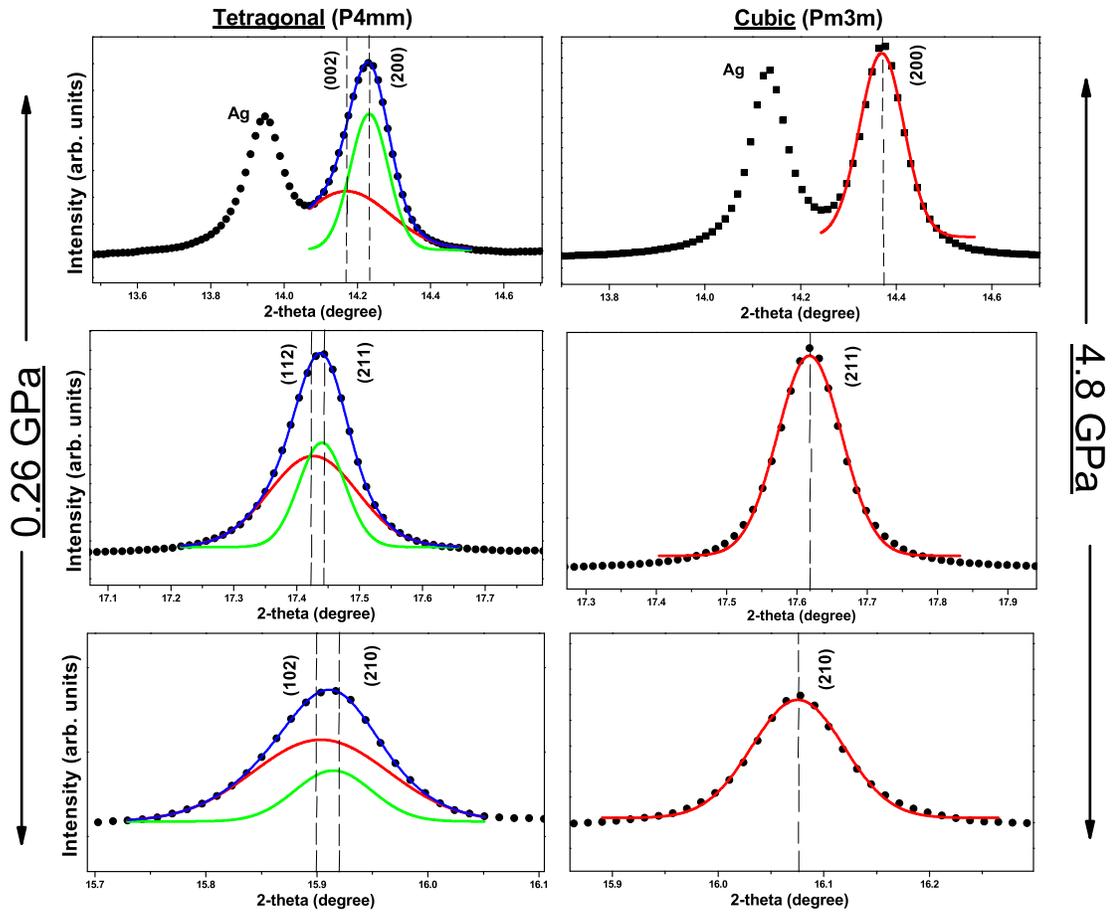}
		\caption{Comparison of Bragg peaks at 0.26 GPa (tetragonal phase) and at 4.7 GPa (cubic phase) respectively. All the doublet peaks merge to single Bragg reflection.}
	%\end{center}
\end{figure}

\begin{figure}
	%\centering
	\includegraphics{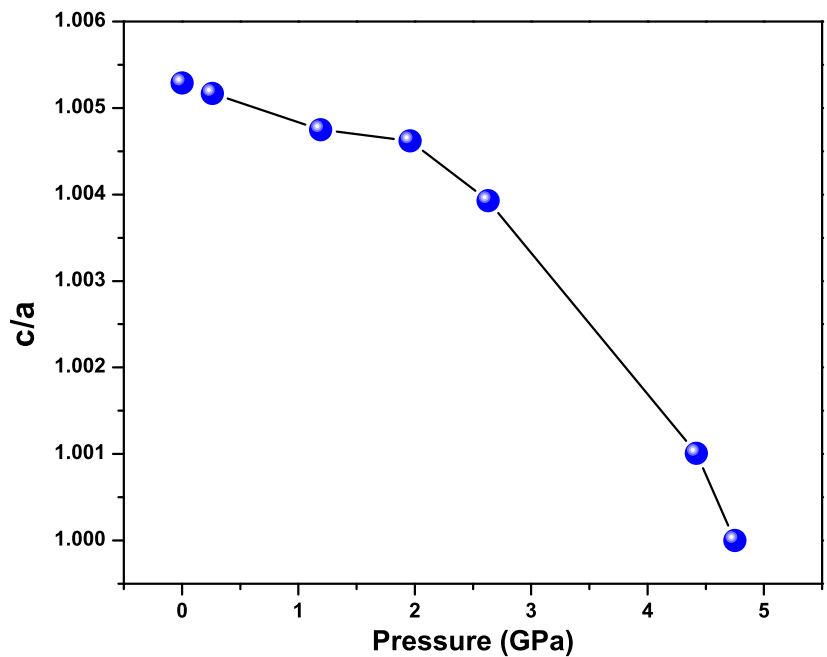}
	\caption{The change in the lattice parameter c/a ratio with pressure.}
\end{figure}

\begin{figure}
	%\centering
	\includegraphics{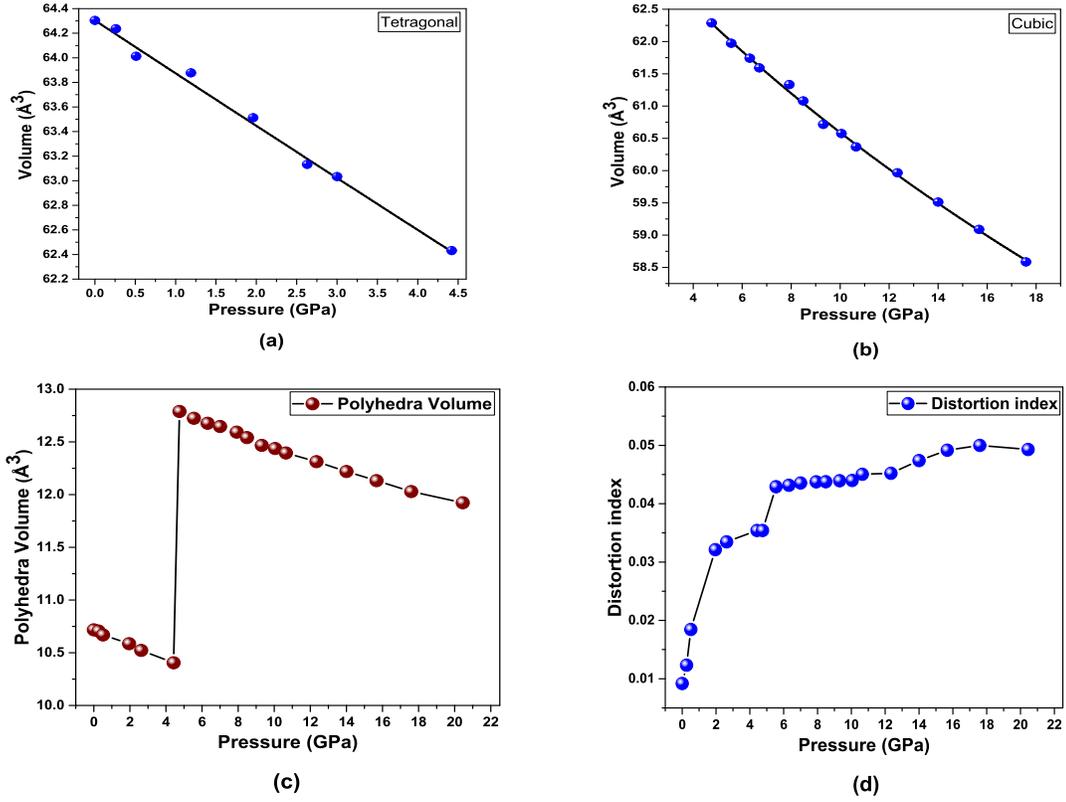}
	\caption{Birch-Murnaghan equation of state fit of pressure-volume data of BCZT in (a) tetragonal phase, (b) and cubic phase. (c) Change in polyhedral volume and (d) distortion index of $(Ti/Zr)O_6$ Octahedra in BCZT is shown. A huge change in octahedra volume is seen at the tetragonal to cubic transition at 4.7 GPa. The distortion index (d) gets saturated after reaching the cubic phase but beyond 10 GPa, followed by a slight variation.}
\end{figure}
\end{document}